\documentclass{aa}
\usepackage{graphicx,natbib}
\usepackage{latexsym}
\newcommand{\mybf}[1]{#1}
\newcommand{\mymbf}[1]{\ensuremath{#1}}

\newcommand{\mynbf}[1]{#1}

\def\aa_{\aap}
\def\aas{\aaps}
\def\plb{Phys.\ Lett.\ B}
\def\cpp{Contrib.\ Plasma Phys.}
\def\rmp{Rev.\ Mod.\ Phys.}

\newcommand{\tot}{\mathrm{d}}
\newcommand{\He}{\element[][]{He}}
\newcommand{\Hen}{\element[++][]{He}}
\newcommand{\citeposs}[1]{\citeauthor{#1}'s \citeyearpar{#1}}
\newcommand{\citealtposs}[1]{\citeauthor{#1}'s \citeyear{#1}}
\newcommand{\rbox}[1]{\raisebox{1.5ex}[0pt]{#1}}
\newcommand{\bref}[1]{\mbox{(\ref{#1})}}
\begin{document}

\title{Microscopic diffusion of partly ionized metals in the Sun and
metal-poor stars} 

\author{H.~Schlattl}

\institute{
Astrophysics Research Institute, Liverpool John Moores
University, Twelve Quays House, Egerton Wharf, Birkenhead CH41 1LD,
United Kingdom}

\offprints{Helmut Schlattl, \\ \email{hs@astro.livjm.ac.uk}}

\date{Received 16 April 2002 / Accepted 21 August 2002}

\titlerunning{Microscopic diffusion of partly ionized metals}

\abstract{An improved microscopic diffusion in stars is presented
considering in detail the partly ionized stages of metals. Besides,
the influence 
of degenerate electron-gas and of the contribution of radiation to the
total pressure has been accounted for. The solution of the diffusion equations
is then performed following the scheme of \cite{Diffc}. By defining one mean
charged ion per element very few modifications are necessary to solve
the improved diffusion scheme. 
(A portable FORTRAN routine is provided.)
The change in the sound-speed profile of a solar model obtained with
the new diffusion description is at most about 25\% at $r=0.6
\,R_\odot$. The biggest effect on low-mass stars is expected near the
turn-off, where the convective envelope is shallowest. However, only a
difference of at most 40~K in the effective temperature could be
observed when assuming either fully or 
partly ionized metals in the diffusion equation. Nevertheless, the
surface metal distribution is strongly altered.

}

\maketitle

\section{Introduction} 
In the last few years the precise measurement of solar
oscillations has challenged theorists to compute solar models of
gradually higher accuracy. This demanded an
improvement of the existing input physics like equation of state
\citep{OPEOS}, opacities \citep{OP96} and nuclear reaction rates
\citep{Adel98}. In 
addition, the formerly neglected process of microscopic diffusion has
been found to improve the agreement of solar models with
helioseismic data considerably \citep{BP92}. Nevertheless, there 
exists a significant discrepancy in the sound speed just below the
convective between the theoretical predicted and the helioseismic
determined value, the reason of which is still unknown.

\mybf{Unfortunately, the expression ``diffusion'' is not always used in the
literature to cover the same physical process(es). In order to
avoid confusion, 
``diffusion'' is defined in this work like, e.g., in
\cite{Bah98} or \cite{SGW00}, to describe 
abundance changes due to pressure, temperature, and concentration
gradients neglecting the effects of radiative forces.} 

\mybf{Beside solar
models} diffusion is now also implemented
in many stellar calculations, e.g., 
for the computation of globular cluster isochrones
\citep{CCIF97,SGW00}. With the improved models globular clusters have
been found to be about \mbox{1\,Gyr} younger than determined previously
by models without diffusion.

However, recent observations of the surface iron abundance of near
turn-off stars \citep{RCBB01,RC02} suggests that diffusion is much less
efficient 
in metal-poor stars with thin convective envelopes than theoretically
predicted. Even worse, the models predict an almost total depletion of
heavy elements in the surface of such stars at the turn off (see
Sect.~\ref{lowmass}). Including the \mynbf{here
neglected} effect of
radiative forces, the opposite effect can be obtained for some elements, e.g.,
\element[][]{Fe} may then be 
enhanced considerably at the surface \citep[]{RMR02}. \mynbf{Thus,
radiative levitation is important in certain evolutionary phases, and
should be included in future models.} 

\mynbf{Nevertheless,} by assuming an additional mixing process below
the convective envelope 
it would be possible to inhibit any diffusion process. Such a process
is also favoured to reduce the discrepancy in the sound speed
just below the solar convective zone \citep{Richard96}.

Various mechanisms have been proposed to cause additional
mixing. Before, however, being able to determine the extent of
additional mixing processes or other non-standard physics, diffusion
should be followed accurately. Therefore, the validity
of physical assumptions used to compute the diffusion efficiency is
investigated in this work.  
A common assumption in the calculation of diffusion constants is the
complete ionization of all elements. Basically in all stellar model
computations including diffusion \citep[e.g.][]{CCIF97,SGW00,WS00,CFN01} this
approximation is made. \mybf{Exception are the models of \cite{RMT00} or
\cite{RMR02} which account for partial ionization and radiative
levitation in the diffusion treatment.}

Since the
cross section for the main scattering process of ions in the stellar plasma,
the Coulomb scattering, is proportional to the square of the ion
charge, deviations from complete ionization may have an important
influence on the diffusivity of the elements. But the diffusion
constant of a specific element is not simply direct proportional to
the cross section, because diffusion has to obey the laws of mass
and charge conservation. Thus, the cross section of each element
has to be considered in relation to the ones of all other elements.
An exact knowledge of the ionization stage of each element
is therefore necessary.

In order to obtain more accurate microscopic diffusion constants the
assumption of fully ionized metals is dropped in this work. Instead,
the ionization stage of each metal is considered in detail, where
the ionization degrees of each element are determined by using an
up-to-date EOS of \cite{Irwin}. 

In the next section the implementation of partly ionized
elements into the solution of Burgers' equations~\citep{Burgers} is
described. The changes in the solar sound-speed profile and in the
evolution of metal-poor stars using this improved description are
discussed afterwards.  

\section{Partly ionized elements in Burgers' equation}
The method to calculate diffusion constants as
described by \citet{Diffc}, denoted TBL hereafter, is followed, where
the basic concepts and quantities can be found. The aim is to improve this
method in particular by including partly ionized atoms. In addition,
a possible non-ideal electron gas and the
contribution of radiation pressure to the hydrostatic-equilibrium
equation are taken into account\footnote{A FORTRAN77 routine containing
all the non-standard effects in the computation of the diffusion
constants is publicly accessible under
\texttt{http://www.astro.livjm.ac.uk/$\mathtt{\tilde{~}}$hs}.}, the
effects of which are neither 
considered by TBL.  
Besides, in this section nuclear burning is neglected; it can simply  
be added afterwards\footnote{In the models presented nuclear burning is, of
course, fully included.}. 

\subsection{Basic equations}
An overall neutral plasma consisting of $S$
species is assumed, where a species is defined to be an element in one particular  
ionization stage. Electrons are supposed to be species
No.~$S$, and are usually not mentioned explicitly. Burgers'
equation for mass conservation can then be 
written as (cf.~Eq.~(11) in TBL)
\begin{equation}\label{mass}
\frac{\partial n_s}{\partial
t}=-\frac{1}{r^2}\frac{\partial}{\partial r}\left(r^2 \omega_s n_s
\right),
\end{equation}
where $r$ is the radial distance from the centre, and $n_s$ is the
number density of ion $s$. 
The diffusion 
velocity $\omega_s$ has to be determined by solving the momentum- and
energy-conservation equations \citep[cf.~Eqs.~(18.1) and (18.2) of][neglecting magnetic fields]{Burgers}
\begin{equation}\label{opress}
\frac{\tot P_s}{\tot r} + \frac{\rho_s}{\rho}
\frac{\tot P_\mathrm{gas}}{\tot r}  - \rho_{es} E =
f_1(\tens{K}, \vec{m}, \vec{\omega}, \vec{r_h})
\end{equation}
and
\begin{equation}\label{temp}
\frac{5}{2}P_s\frac{\tot \ln T}{\tot r} = f_2(\tens{K}, \vec{m}, \vec{\omega}, \vec{r_h}),
\end{equation}
where $P_s$, $\rho_s$($=n_s m_s$), and $\rho_{es}$($=n_s q_s$)
are the partial pressure, mass density, and charge density,
respectively, for ion 
$s$ with $m_s$ being its mass and $q_s$ 
its charge; $E$ is the electric field strength. The vectors $\vec{m}$
and $\vec{\omega}$ contain the mass and diffusion velocity for each
\mbox{ion $s$}. For a detailed definition of the friction-coefficient tensor
$\tens{K}$, the residual heat-flow vector 
$\vec{r_h}$, and the functions $f_1$ and $f_2$ the reader is referred to
TBL.

The total pressure $P$ is composed of radiation
($P_\mathrm{rad}$) and gas pressure ($P_\mathrm{gas}$), thus
\[
P = P_\mathrm{rad} + P_\mathrm{gas} = P_\mathrm{rad} + \sum P_s = \sum \frac{1}{\beta} P_s,
\]
where the standard definition $\beta =
P_\mathrm{gas}/P$ has been employed. Since 
the departure from local charge neutrality is very small (cf.~TBL),
$\rho_e E$ is close to zero, and thus 
the usual hydrostatic-equilibrium equation 
\begin{equation}\label{hystat}
\frac{\tot P}{\tot r} = -g \rho
\end{equation}
can be used, with $g$ being the gravitational acceleration. Hence,
Eq.~\bref{opress} becomes
\begin{equation}\label{press}
\frac{\tot P_s}{\tot r} + \beta g \rho_s - \frac{\rho_s}{\rho}
\frac{\tot \beta}{\tot r}  - \rho_{es} E =
f_1(\tens{K}, \vec{m}, \vec{\omega}, \vec{r_h})
\end{equation} 

By two additional constraints, current neutrality,
\begin{equation} \label{neutcur}
\sum q_s n_s \omega_s = 0,
\end{equation}
and local mass conservation,
\begin{equation} \label{localmass}
\sum m_s n_s \omega_s = 0,
\end{equation}
Eqs.~\bref{temp} and \bref{press} are completed to form a closed set of
$2S+2$ equations for the unknown quantities
$\vec{\omega}$, $\vec{r_h}$, $g$ and $E$.

For a large number of species with varying abundances
these equations are preferentially solved
numerically. For this purpose, they are reformulated to be more
suited for stellar evolution calculations. 

The partial pressure $P_s$ for atom $s$ is given by the ideal gas
equation 
\begin{equation}\label{ideal}
P_s = n_s k_{\mathrm{B}} T.
\end{equation}
The electron pressure in
stellar envelopes is well described by this formula, too.
However, in the deep radiative regions of low-mass stars, and in
particular in white dwarfs, the electron gas is degenerate. 
Therefore, an ``effective'' electron density is introduced
\begin{equation}\label{defne}
\tilde{n}_e = \frac{P_e}{k_{\mathrm{B}}T}
\end{equation}
such that the total pressure can be written 
\[
P = \frac{1}{\beta}\sum n_s k_{\mathrm{B}} T,
\]
where $n_S = \tilde{n}_e$. Furthermore the
concentration $C_s$ is defined by
\begin{equation}\label{defc}
C_s = \frac{n_s}{\tilde{n}_e},
\end{equation}
analogously to TBL.

With $C = \sum C_s$ one can now write
\begin{equation}
\frac{P_s}{\beta P} = \frac{P_s}{\sum{P_s}} = \frac{n_s}{\sum{n_s}} =
\frac{C_s}{C},
\end{equation}
and hence the first term of Eq.~\bref{press} becomes
\begin{equation}\label{psequ}
\frac{\tot P_s}{\tot r} = \frac{\tot \frac{C_s}{C}\beta P}{\tot r} = 
\frac{C_s}{C}\beta P \left(\frac{\tot \ln \beta P}{\tot r} + \frac{\tot \ln
(C_s/C)}{\tot r} \right).
\end{equation}
Taking into consideration that in TBL the pressure $P$ has been
defined to be solely the gas pressure, this term seems to agree with
TBL. However, here $C_s$ has been defined via 
Eq.~\bref{defc} instead of TBL's Eq.~(22) ($C_s = n_s/n_e$), which
accounts for a  possible degeneracy of the electron gas.

Demanding charge neutrality, $\sum_{s \neq e} q_s n_s = -q_e n_e$, one
gets rid of the dependence on one species, e.g.~$\Hen$. Thus,
\begin{equation}\label{neutc}
\zeta_{\Hen} C_{\Hen} = (\tilde{C}_e - \sum_{s\neq\Hen,e}
\zeta_s C_s), 
\end{equation}
where $\zeta_s$ is the ionization stage of ion $s$ ($\zeta_s$
= 0 for neutral atoms) and 
\begin{equation}\label{defce}
\tilde{C}_e = \frac{n_e}{\tilde{n}_e}.
\end{equation}

Similarly, the l.h.s.~of Eq.~\bref{temp} becomes
\begin{equation}\label{tfunc}
\frac{5}{2}P_s\frac{\tot \ln T}{\tot r} =
\frac{5}{2}\frac{C_s}{C}\beta P\frac{\tot \ln T}{\tot r}.
\end{equation}

Setting Eqs.~\bref{ideal}--\bref{tfunc} into
Eqs.~\bref{temp}--\bref{localmass} one finds, after some algebraic
manipulations, that the diffusion velocity can be written 
in general as (see Appendix~\ref{appx})
\begin{equation}
\omega_s  =  \frac{T^{5/2}}{\rho}\xi_s,
\end{equation}
where 
\begin{eqnarray} \label{diffdef}
\xi_s & = & A^P_s\frac{\tot \ln P}{\tot r}
+ A^T_s\frac{\tot \ln T}{\tot r} + \sum_{t \neq e,\Hen} A^C_{t,s}\frac{\tot \ln C_t}{\tot r}
\nonumber \\  
&& +  A^\beta_s\frac{\tot \ln \beta}{\tot r}
+ A^C_{e,s}\frac{\tot \ln \tilde{C}_e}{\tot r} + \sum_{t \neq
e} A^\zeta_{t,s}\frac{\tot \ln \zeta_t}{\tot r}. 
\end{eqnarray}
The diffusion coefficients $A^P$, $A^T$, $A^C$, $A^\beta$, and $A^\zeta$ are
obtained by the solution of
Eqs.~\bref{temp}--\bref{localmass}. \mybf{For convenience, the six terms of
Eq.~\bref{diffdef} are  
denoted $\mymbf{\xi_s^P}$, $\mymbf{\xi_s^T}$, $\mymbf{\xi_s^C}$, 
$\mymbf{\xi_s^\beta}$, $\mymbf{\xi_s^e}$,
and $\mymbf{\xi_s^\zeta}$, respectively.} Compared to the analogous equation
in TBL (Eq.~(40)), three
additional terms appear here. \mynbf{They should not be considered as
new types of diffusion but rather as corrections to pressure (term 4)
and concentration diffusion (terms 5 and 6)}.

\mynbf{In particular,} the fourth term in Eq.~\bref{diffdef}
\mybf{($\mymbf{\xi_s^\beta}$)}
corrects for the appearance of a radiation pressure in the
hydrostatic-equilibrium equation. The fifth term \mybf{($\mymbf{\xi_s^e}$)}
takes into account that for degenerate electrons
the ideal linear relation between pressure and density is not 
valid any more (Eq.~\bref{defce}). Thus, this term is only important,
when strong gradients in the electron degeneracy appear (e.g.~in white
dwarfs).

\begin{figure}
\resizebox{\hsize}{!}{\includegraphics{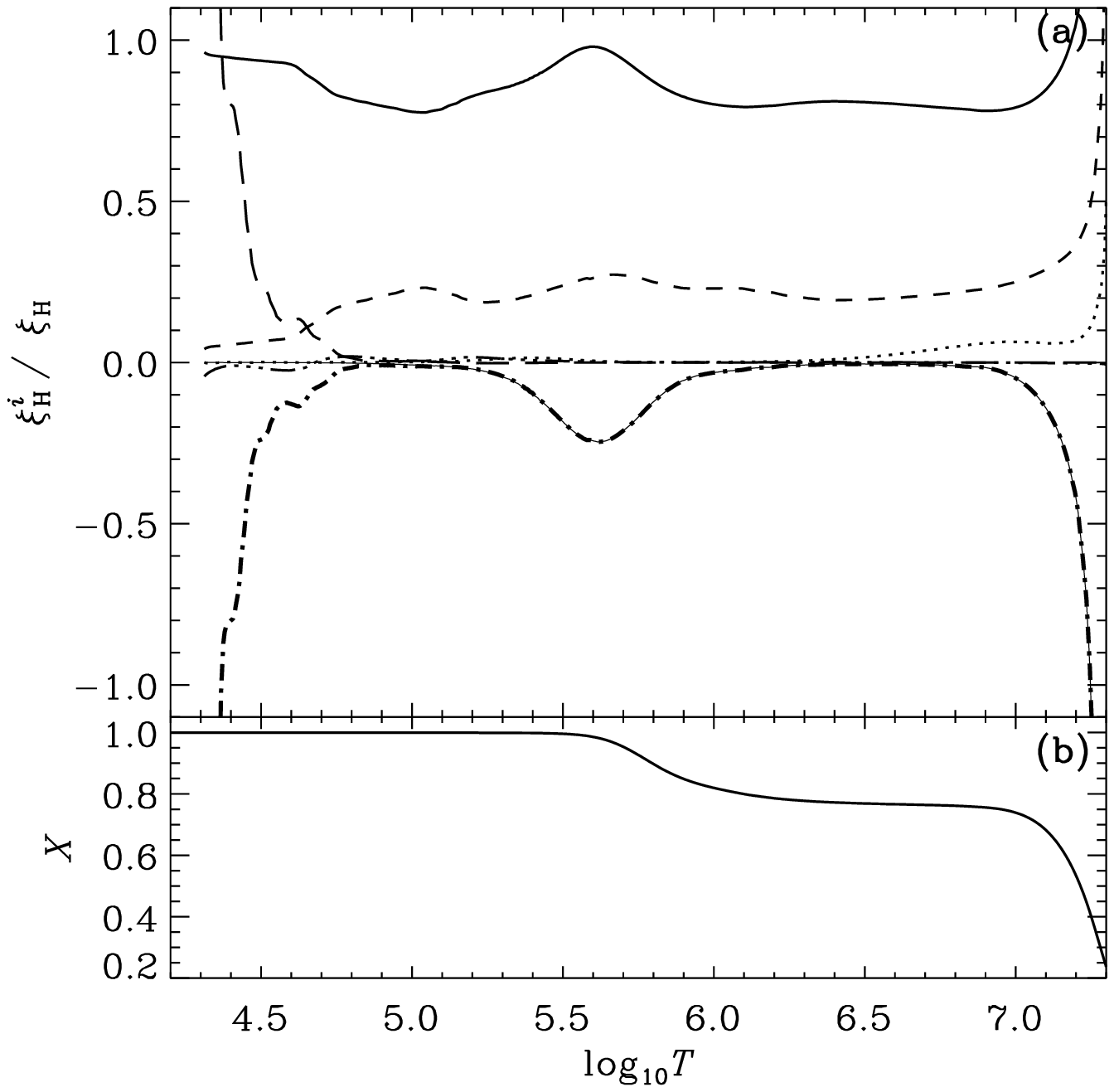}}
\caption{\textbf{a)} \mybf{The contributions of the six terms in
Eq.~\bref{diffdef} to 
the diffusion velocity of hydrogen in a $\mymbf{1.1~M_{\odot}}$ star
near the TO with initial
$\mymbf{\mathrm{[\element[][]{Fe}/\element[][]{H}]=-1.3}}$:
$\mymbf{\xi_s^P}$ (solid line), $\mymbf{\xi_s^T}$ (short-dashed),
$\mymbf{\xi_s^C}$ (dash-dotted), $\mymbf{10\times\xi_s^\beta}$
(dash-dot-dot-dotted), $\mymbf{10\times\xi_s^e}$ (dotted), and
$\mymbf{\xi_s^\zeta}$ (long-dashed). The thin solid line represents
$\mymbf{\xi_s^C+\xi_s^\zeta}$. \mynbf{\textbf{b)} The hydrogen mass fraction
in this model.}} 
\label{diffv}}
\end{figure}

\mybf{In Fig.~\ref{diffv}a the contribution of each of the 6 terms in
Eq.~\bref{diffdef} to the diffusion velocity $\mymbf{\omega}$ of
\element[][]{H} in a $\mymbf{1.1~M_{\odot}}$ star near the turn off (TO) are
shown. Pressure ($\mymbf{\xi_s^P}$) and temperature diffusion
($\mymbf{\xi_s^T}$) are clearly dominating in most parts of the
stellar interior. Near the centre, where already most of the H has
been consumed, 
concentration diffusion ($\mymbf{\xi_s^C}$) becomes important. The
contribution of $\mymbf{\xi_s^\beta}$ remains very small throughout
the star, while $\mymbf{\xi_s^e}$ increases toward the centre, where
the electron degeneracy is gradually increasing.}

\mybf{The last term in Eq.~\bref{diffdef} ($\mymbf{\xi_s^\zeta}$),
which originates from the demand for charge neutrality (Eq.~\bref{neutc}),
needs a more thorough consideration. It would disappear,
if each ion is treated as a separate species, because the ionization
stage $\zeta$ of each species, i.e., the number of
electrons per ion, is constant. Charge neutrality is
then, as in the case of complete ionization, automatically conserved
by the definition of $A^C$. 
But if, as shown below, a ``mean'' ion per element is defined which is
in an average ionization stage $\overline{\zeta}$, the number of
electrons per species is varying in the ionizing regions of a star.
This would cause artificial concentration
diffusion. Due to $\mymbf{\xi_s^\zeta}$ this
effect is compensated, ensuring that only the total
number density of a certain element is relevant for its concentration
diffusion, independent of the respective ionization stages in two
neighbouring layers.}

\mynbf{Note that using one mean charged ion per element cannot lead to
ambipolar diffusion \citep[see][and references therein]{BM91}. 
Hence the ionization state of each element is determined in this
case solely by the local thermodynamic 
equilibrium and cannot be modified by diffusion.}

\begin{figure}
\resizebox{\hsize}{!}{\includegraphics{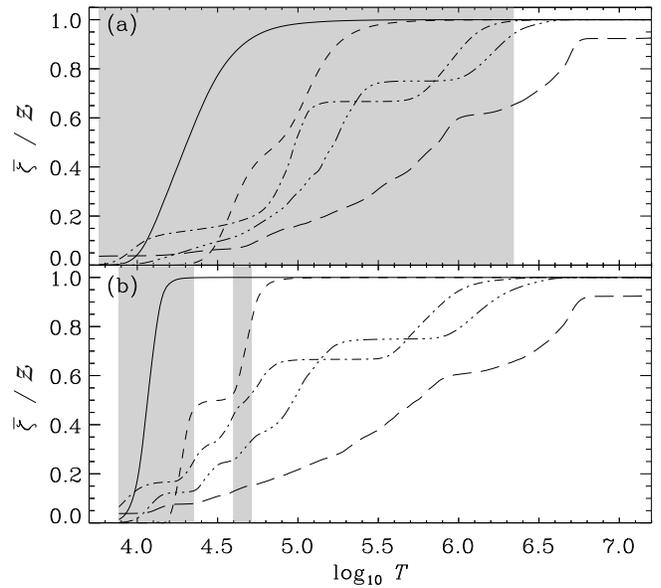}}
\caption{\textbf{a)} The mean ionization degrees $\overline\zeta$ 
of \element[][]{H} (solid line), \He~(short dashed), \element[][]{C}
(dash-dotted), \element[][]{O} (dash-dot-dot-dotted), and
\element[][]{Fe} (long dashed) in the Sun, normalized to their 
respective charge number $\mathcal{Z}$ (obtained from \citealtposs{Irwin}
EOS).  The
envelope convective zone is located within the grey
shaded area. \textbf{b)} Same
as a) but for a $1.1~M_{\odot}$ star near the TO with
$\mathrm{[\element[][]{Fe}/\element[][]{H}]=-1.3}$.
\label{iondeg}}
\end{figure}

\mybf{An example is provided in Fig.~\ref{diffv}a, where 
mean charged ions per element were used to compute the contributions to
the diffusion velocity. In 
the layers near $\mymbf{\log_{10} T\approx 4.4}$, just below the
convective envelope (c.f.~Fig.~\ref{iondeg}), \element[][]{H} is not completely ionized
($\mymbf{\zeta_\mathrm{H}\approx 0.995}$). This small
deviation from complete ionization causes a large negative contribution 
of the concentration diffusion term (dash-dotted line) to
$\mymbf{\omega_\mathrm{H}}$, although the hydrogen abundance is
constant in this region (Fig.~\ref{diffv}b). However, by the last term in
Eq.~\bref{diffdef} this artificial concentration diffusion is
counterbalanced such that the sum of $\mymbf{\xi_s^C}$ and
$\mymbf{\xi_s^\zeta}$ remains zero in this area (thin solid line in
Fig.~\ref{diffv}a)}.

The diffusion of ions into regions, where they are
out of thermodynamic equilibrium,  yields an additional energy
sink (or source). However, the ionization of the most abundant elements,
which are usually hydrogen and helium, leads in most relevant cases to the
development of convective zones (Fig.~\ref{iondeg}), where mixing
occurs on much shorter timescales than diffusion. Thus, in the
radiative layers only less abundant elements may be in differently
ionized stages. The out-of-equilibrium distribution of their
ions leads only to a small amount of energy compared to the total
internal energy of the plasma, and is thus neglected here.

\subsection{Linear approximation for $\omega_s$}

Using a mean ionization stage for all elements, instead of
treating each ion separately, is valid if the diffusion 
velocity $\omega_s^i$ is a linear function of the charge of ion $i$ of
element $s$. Then
\begin{equation}\label{lindiff}
\omega_s = \omega_s(\zeta) = \omega_{s,0} + \zeta\,  \delta \omega_s.
\end{equation}
and therefore
\begin{equation}\label{linom}
\sum_i\omega^i_s n^i_s = \sum_i(\omega_{s,0} + \zeta_s^i\,\delta
\omega_s)n_s^i = (\omega_{s,0} + \overline{\zeta}_s \,\delta
\omega_s) n_s
\end{equation}
is obtained, where $\zeta_s^i$ are the discrete ionization stages of $s$, and
$\overline{\zeta_s}$ is the mean charge defined by
$\overline{\zeta_s} = \left(\sum_i \zeta_s^i n_s^i\right)/n_s$ with 
$n_s = \sum_i n_s^i$. 

In order to verify to what extent the assumption of $\omega_s$
being linear in $\zeta$ is justified, the diffusion
coefficients for a simple mixture of hydrogen, helium and iron with
$X=0.76$, $Y=0.23$ and $X(\mathrm{Fe}) = 0.01$ have been computed.
Conditions similar to those
at the base of the solar convective envelope were chosen, i.e.,
$T=2.24\times 10^6~\mathrm{K}$ and $\rho = 0.19~\mathrm{g/cm^3}$.

\begin{figure}
\resizebox{\hsize}{!}{\includegraphics{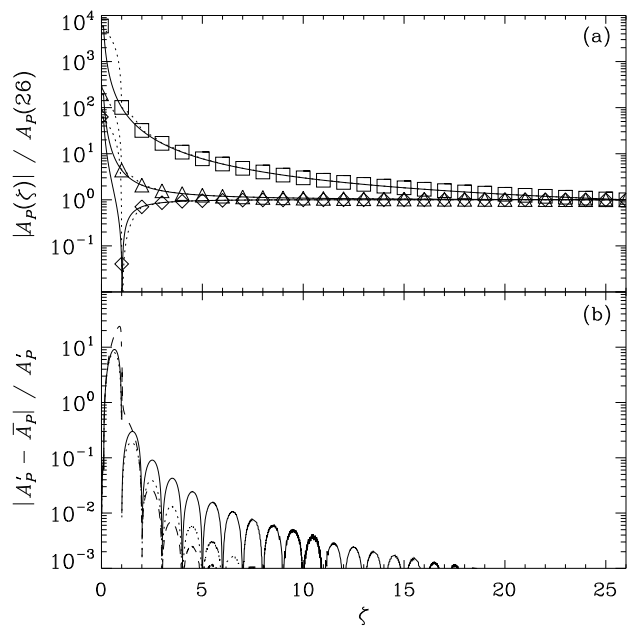}}
\caption{\textbf{a)} Dependence of the pressure diffusion constant $A_P$
of \element[][]{H} ($\triangle$), \He~($\Diamond$) and
\element[][]{Fe} ($\Box$) on the charge 
($\zeta$) of the \element[][]{Fe}-ion in a mixture of $X=0.76$, $Y=0.23$ and 
$X(\mathrm{Fe}) = 0.01$ at $T=2.24\times 10^6~\mathrm{K}$ and $\rho =
0.19~\mathrm{g/cm^3}$. Using a mixture of two different
\element[][]{Fe}-ions with 
charges $\zeta(i)$ and $\zeta(i+1)$ the dotted line is obtained for the mean
diffusion 
constants $\overline{A}_P$ (``exact value''), while the solid line
represents the result employing the mean charge $\overline{\zeta}$ (``linear
approximation''). At $\zeta \approx 1$ $A^\prime_P(\mathrm{He})$
changes its sign. \textbf{b)} The relative difference between $\overline{A}_P$ and
$A^\prime_P$ for 
\element[][]{H} (dotted line), \He~(dashed) and \element[][]{Fe}
(solid).\label{feion}}
\end{figure}

In Fig.~\ref{feion}a the symbols mark the pressure
diffusion coefficients $A_P$ of \element[][]{H}, \He, and
\element[][]{Fe} for different ionization 
levels of the latter, while \element[][]{H} and \He~are assumed to be fully
ionized. For \ion{Fe}{I} the interactions 
can no longer be described by Coulomb scattering, but are due to the
atomic polarisability~\citep[see e.g.][]{MMR78}. It has been found 
that the diffusion coefficients of a neutral atom is about a factor of
100 higher than of its single charged ion~\citep[see][]{GLAM95}. To
obtain a similar 
value but using the Coulomb scattering description an 
``effective'' ionization degree of 0.1 for neutral atoms is adopted. This
certainly only yields approximate values for the diffusion constants,
and should be improved in future works.

Obviously, $A_P$ is not a simple linear function of $\zeta$, but
with increasing ionization the non-linearity becomes smaller.
The dotted line in Fig.~\ref{feion}a shows the mean diffusion
coefficient $\overline{A}_P$ for a mixture consisting of two Fe-ions with
number density $n_i$ and $n_{i+1}$ and discrete charges $\zeta_i$ and
$\zeta_{i+1}$, respectively. If instead the mean  
ionization stage $\overline{\zeta} = \frac{n_i\zeta(i) +
n_{i+1}\zeta(i+1)}{n_i + n_{i+1}}$ is used, $A^\prime_P(\zeta)$ follows the
solid line.  

\begin{figure}
\resizebox{\hsize}{!}{\includegraphics{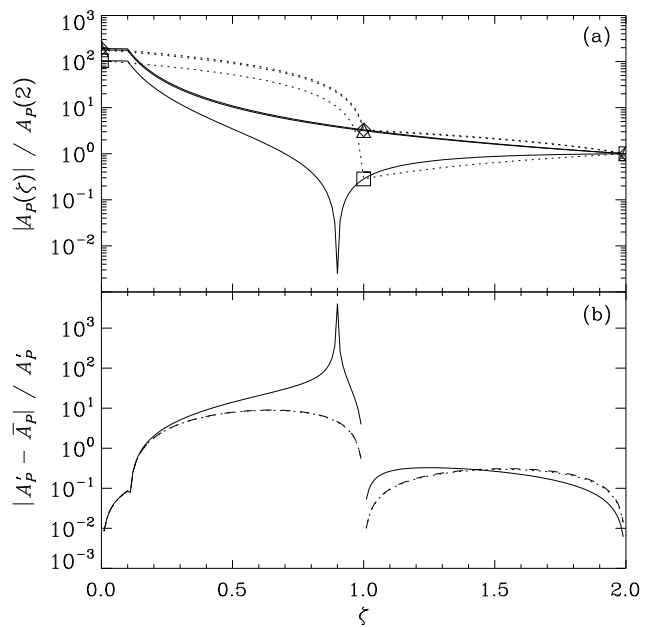}}
\caption{Same as Fig.~\ref{feion} but for varying \He-ionization
degrees. Note that at $\zeta \approx 0.9$
$A^\prime(\element[][]{Fe})$ changes the sign.\label{heion}}
\end{figure}

In Fig.~\ref{feion}b the relative error between the two
values is plotted. Because of the large mean free path of
\ion{Fe}{I}, $A_P(\He)$ has a different sign for neutral than for
ionized \element[][]{Fe} causing \He~to diffuse in the
same direction as \element[][]{H}. This behaviour can also be
observed, if \element[][]{Fe} is assumed to be fully ionized,
and the ionization stage of \He~is varied (Fig.~\ref{heion}).
Then \element[][]{Fe} would diffuse in the same direction as
\element[][]{H}, when mostly \ion{He}{I} is present in the plasma.

But in real stars such "anomalous" diffusion
would rarely occur, as the ionization of hydrogen and helium leads to
the development of convective regions, where mixing is much faster
than diffusion. In these 
regions metals are, at least partially, ionized, too
(Fig.~\ref{iondeg}). Therefore, in the diffusion-dominated radiative
layers helium and metals almost never are in their lowest ionization
stages, and thus the difference between $\overline{A}_P$ and
$A^\prime_P$ remains small.

In some cases, however, the \element[][]{He} abundances at the outer
layers are reduced considerably by diffusion such that no convection
develops due to \element[][]{He} ionization. Then the
diffusion of each \He-ion should be treated separately. But taking
into consideration that the diffusion velocity of \ion{He}{I} is not
determined very accurately and the \He\ abundance in that case is very
low, using the mean ionization degree of 
\element[][]{He} may still yield sufficiently accurate results.

In summary, the diffusion of partly ionized elements in stars can in
most relevant cases be determined by using one ``effective'' ion for
each species carrying the mean ionization degree $\overline{\zeta}$. With
this simplification the computation of the diffusion of partially
ionized elements is not connected with larger computational time than
using fully ionized particles and remains therefore feasible.

\section{Stellar and solar models}
For the calculation of the stellar models a descendant of the
program by \cite{KWH} in the latest version described in \cite{S00} is
used.
In particular the opacities tables of \cite{OP96} and the
nuclear reaction rates given by \cite{Adel98} are employed. While
previously the 
equation-of-state tables of \cite{OPEOS}, known as OPAL- or ACTEX-EOS,
has been implemented, now 
the analytic description of \cite{Irwin} is applied. This step has
become necessary, as the \mybf{publicly accessible} 
\citeposs{OPEOS} tables do not provide quantities like the electron
density or ionization degrees. \citeposs{Irwin} equation of state
(IRWIN-EOS) is a further development of the SIREFF-EOS, which itself
was built from the EFF-EOS \citep{EFF}. While EFF could
be solved during the stellar evolution program (``in-line''), 
IRWIN is pretty slow, so that I produced spline tables on a
pressure-temperature grid for different hydrogen and metal 
fractions. The metal distribution was chosen from \cite{GN93}. 
 
The chemical evolution is determined by solving
simultaneously nuclear network and diffusion equation in a
common scheme~\citep{PhD}, where convection is treated as a fast
diffusive process. We follow the abundance changes of \element[][]{H},
\element[][3]He, \element[][4]He, \element[][12]{C}, \element[][13]{C},
\element[][14]{N}, \element[][15]{N}, \element[][16]{O},
\element[][17]{O}, \element[][20]{Ne}, \element[][24]{Mg},
and \element[][56]{Fe}, which implies that 92\% of
the total metallic mass is considered. 

\subsection{Solar models}

In my solar models the outer layers above an optical depth of 1000
have been taken from a 2D-hydrodynamical atmosphere model
of~\cite{2dhydro}, which is determining also the outer 
boundary condition of the interior solar model. These atmospheres
extend down to regions, where the stratification is almost adiabatic,
including the superadiabatic layers just below the surface. By this means
improved p-mode frequencies could be obtained like with formerly
applied 1D-atmospheres~\citep{SWL97}, and the
models become almost independent of the convection theory applied. For
the latter the description of \cite{CM2} is used.
As usual, the convection parameter $\alpha_\mathrm{CM}$, initial
helium and metal 
content are adjusted to yield models with present solar luminosity, radius
and metal fraction $Z/X$. In order to explore the effect of different
diffusion approaches these solar constraints are reproduced with
an accuracy better than $10^{-6}$. A summary of all solar models computed in 
this work is provided in Table~\ref{soltab}. 

\begin{table}
\caption{Physics in solar models\label{soltab}}
\begin{tabular}{l|c|ccc} \hline \hline
& & \multicolumn{3}{c}{Effects included in diffusion} \\
\rbox{Model} & \rbox{EOS} & non-ideal $e^-$  & $P_\mathrm{rad}$ & partly
ionized \\ \hline
OP1 & OPAL96 & --- & --- & ---\\
OP2 & OPAL01 & --- & --- & ---\\
IR1 & IRWIN & --- & --- & ---\\
IR2 & IRWIN & X & --- & --- \\
IR3 & IRWIN & X & X & ---\\
IR4 & IRWIN & X & X & X \\ \hline
\end{tabular}
\end{table}

In a first step models using the diffusion
description of TBL containing OPAL- or IRWIN-EOS have been produced. Recently, an
updated version of the OPAL-EOS has become available (denoted OPAL01,
hereafter). The main differences compared to the version of 1996
(OPAL96) are the improvement of the activity expansion method for
repulsive interactions~\citep{Rog01}, the inclusion of
\element[+][]{HeH} and \element[++][]{He} in addition to 
H$_2$, H$^+_2$ and \element[-][]{H}, and a relativistic treatment of
the electrons.  

\begin{figure}
\resizebox{\hsize}{!}{\includegraphics{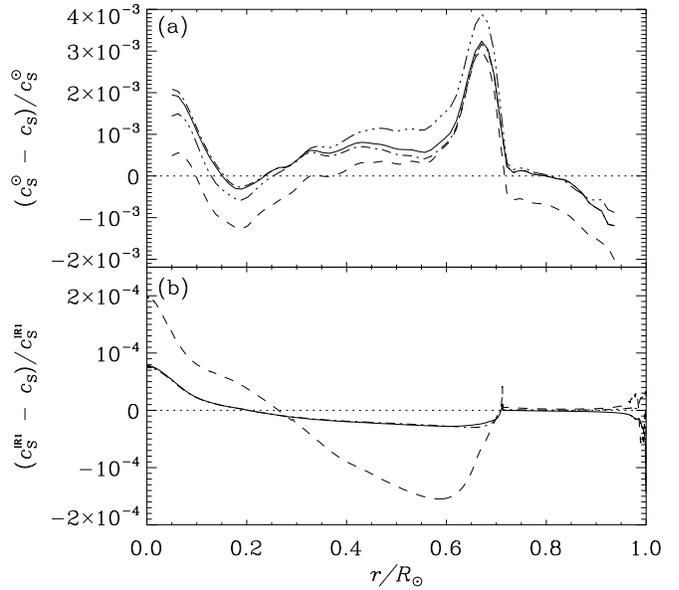}}
\caption{\textbf{a)} Difference in the run of sound speed between 
the seismic model of \protect\citet{BCC97} and solar models OP1
(dashed line), OP2 (dash-dot-dot-dotted), IR1 (solid), and IR4
(dash-dotted), respectively. \textbf{b)}
Sound speed of model IR2 (solid line), IR3 (dash
dotted), and IR4 (dashed) compared to the reference model IR1.
\label{sound}}
\end{figure}

In Fig.~\ref{sound}a the differences of the sound speed between the
seismic model of \cite{BCC97} and 
models with OPAL96-, OPAL01-, and IRWIN-EOS, respectively, are
shown. Interestingly, model OP2 
containing the updated OPAL01-EOS provides a worse
agreement on average with the 
solar sound-speed profile than model OP1 with the older OPAL96-EOS.
The main reason is that the adiabatic
index $\Gamma_1$ decreases when treating electrons
relativistically~\citep{BMP01}. However,
while the overall sound-speed profile has deteriorated,
models with OPAL01-EOS provide a better agreement between the
helioseismologically determined solar age and 
the meteoritic age of the solar system~\citep{BSP02}

Model IR1 was computed with the IRWIN-EOS, where electrons are also
treated relativistically and where H$_2$- and H$^+_2$-molecules
are included. While in the central and convective region ($r>0.7\;
R_{\odot}$), the sound speed of IR1 and OP2 are very similar, between
$0.25 < r/R_{\odot} < 0.6$ IR1 reproduces the solar-sound speed  
profile better than model OP2. It is beyond the scope of this work
to analyze what causes the difference between OPAL01- and IRWIN-EOS. 
Anyway, with the latter the solar sound-speed profile can be
reproduced very well, and it thus provides an at least equally good
description of stellar plasmas as the OPAL01-EOS.

\begin{table*}[t]
\begin{center}
\caption{Characteristic quantities of solar models in this work. The
indices $i$, $s$, $c$, and $\mathrm{C.Z.}$ denote initial, surface,
core and bottom of convective envelope, respectively. 
$\alpha_\mathrm{CM}$  is the parameter of
\protect\citeposs{CM2} convection theory accounting for non-local effects,
and $\Phi_\nu^\mathrm{SK}$ is the total flux of neutrinos produced in the
\element[][8]{B}-decay and the
\element[][3]{He}(p,$e^+\nu_e$)\element[][4]{He}-reaction. In the last two columns the
expected rates for the Gallium (Ga) and Chlorine (Cl) experiments are
provided, where 1 SNU (solar neutrino unit) corresponds to 1 event per
$10^{36}$ target atoms and seconds.\label{solqu}}
\begin{tabular}{l||ccc|*{8}{c}} \hline \hline
Model & $Y_i$ & $Z_i$ & $\alpha_\mathrm{CM}$ & $Y_s$ & $Z_s$ &
$\frac{R_\mathrm{C.Z.}}{R_\odot}$ & $\frac{\rho_c}{\mathrm{gcm^{-3}}}$
& $\frac{T_c}{10^7\mathrm{K}}$
 & $\frac{\Phi_\nu^\mathrm{SK}}{10^6\mathrm{cm^{-2}s^{-1}}}$ & $\frac{\mathrm{Ga}}{\mathrm{SNU}}$ &
$\frac{\mathrm{Cl}}{\mathrm{SNU}}$ \\ \hline
OP1 & 0.27478 & 0.01992 & 0.97803 & 0.24429 & 0.01807 & 0.7131 & 152.01 
& 1.5714 & 5.090 & 128.6 & 7.624 \\
OP2 & 0.27520 & 0.01987 & 1.0264 & 0.24509 & 0.01805 & 0.7132 & 152.02
& 1.5711 & 5.079 & 128.5 & 7.608 \\ 
IR1 & 0.27430 & 0.01987 & 1.0099 & 0.24450 & 0.01807 & 0.7123 & 151.87
& 1.5708 & 5.063 & 128.4 & 7.585 \\
IR2 & 0.27430 & 0.01987 & 1.0105 & 0.24448 & 0.01807 & 0.7124 & 151.92
& 1.5710 & 5.072 & 128.4 & 7.600 \\
IR3 & 0.27429 & 0.01987 & 1.0105 & 0.24449 & 0.01807 & 0.7124 & 151.93
& 1.5710 & 5.072 & 128.4 & 7.598 \\
IR4 & 0.27468 & 0.01996 & 1.0118 & 0.24482 & 0.01806 & 0.7125 & 151.98
& 1.5716 & 5.108 & 128.7 & 7.647 \\ \hline
\end{tabular}
\end{center}
\end{table*}

Starting with model IR1 a series of solar models have been computed
where  subsequently non-standard effects have been added in
the diffusion equation. The relevant physics included in each model
is summarized in Table~\ref{soltab}. 

\begin{figure}
\resizebox{\hsize}{!}{\includegraphics{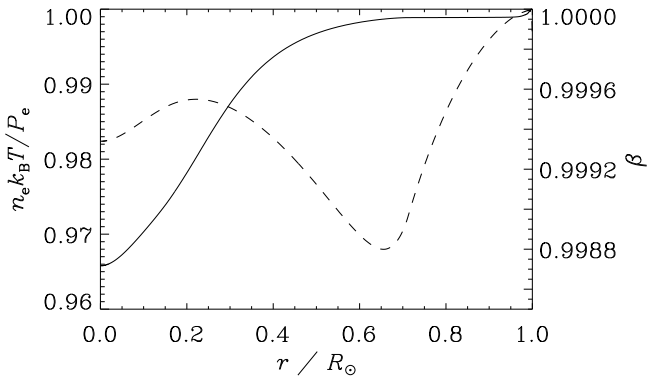}}
\caption{The solid line shows the ratio between the electron
pressure assuming an ideal gas ($n_ek_\mathrm{B}T$) and the actual
electron pressure $P_e$ in 
model IR1, while the dashed line represents $\beta$, the relative
contribution of radiation to total pressure.
\label{eta}}
\end{figure}

Since the electron degeneracy in the Sun is very small
(Fig.~\ref{eta}), its implementation in the diffusion equation alters
the sound speed at most only about 0.01\% in the centre
(Fig~\ref{sound}b), where the degeneracy is highest.

An even smaller change is obtained by taking into account the
radiation pressure which contributes at most about 0.1\% at
$r=0.65\,R_{\odot}$ to the total pressure (Fig.~\ref{eta}). There a
tiny change in the sound speed of about $10^{-5}$ can be observed
(Fig~\ref{sound}b), but overall it remains essentially invariant.

With the inclusion of partly ionized elements in the diffusion
treatment the sound speed has 
increased by almost 0.02\% at $\,r = 0.6\,R_{\odot}$, but decreased by
about 0.01\% in the centre compared to IR3. Taking into account that
the sound speed of IR3 (and equally IR1) has differed by about 0.08\%
from the seismic model for $0.3 < r/R_\odot < 0.65$, an
improvement of about 25\% could be achieved with model IR4. 

In the present stage, however, the discrepancies between
different input physics like equation of state are exceeding this
value and thus the effect of partly ionized elements on diffusion 
appears to be negligible. But, including this effect reduces the
uncertainty in solar models and thus helps to analyze the still existing
shortcoming in model input physics.

\cite{TRM98} has investigated the influence of a detailed treatment of
the ionization stages in the diffusion equation, too. But in
contrast to my results, they found a decrease of the sound speed when
using partly instead of completely ionized elements. However, they
followed a different strategy for computing solar models by using the
same initial $Z/X$ ratio in the models under scrutiny. Furthermore,
they failed to compare directly two
models where the ions are assumed to be either fully or partly
ionized. Instead they compared their model, which contained a detailed
treatment of the ionization stages, with one where the diffusion
velocities of all metals heavier than \element[][]{O} are assumed to
be equal to the velocity of completely ionized Fe. Both
differences might contribute to the discrepancies between our results.

In Table~\ref{solqu} some characteristic quantities of the presented solar
models are summarized. Only very small differences can be observed in
all quantities. With the updated OPAL01- and IRWIN-EOS the core
temperature has been reduced a little compared to model OP1 with the
OPAL96-EOS. However, with the inclusion of partly ionized metals in
the diffusion equation the temperature increases weakly,
yielding a somewhat higher value for IR4 as for OP1, and thus slightly
enhanced neutrino rates.  

\begin{table}
\tabcolsep.55em
\caption{Abundance variations of solar models of this
work. The central abundance of \element[][12]{C} was in all models
0.7\% of the initial one.\label{abunsol}}
\begin{tabular}{l|rrrr|rrr} \hline \hline
& \multicolumn{4}{c|}{Surface (\%)} & \multicolumn{3}{c}{Center (\%)} \\
\rbox{Model} & \multicolumn{1}{c}{\element[][4]{He}} &
\multicolumn{1}{c}{\element[][12]{C}} &
\multicolumn{1}{c}{\element[][16]{O}} & \multicolumn{1}{c|}{$Z$} & 
\multicolumn{1}{c}{\element[][4]{He}} & \multicolumn{1}{c}{\element[][16]{O}} &
\multicolumn{1}{c}{$Z$} \\ \hline
OP1 & $-$11.1 & $-$10.3 & $-$10.1 & $-$10.1 & 133.1  & $-$2.2 & 8.0 \\ 
OP2 & $-$10.9 & $-$10.1 & $-$9.9  & $-$9.9  & 132.7  & $-$2.3 & 7.9 \\
IR1 & $-$10.9 & $-$10.1 & $-$9.9 &  $-$9.9  & 133.3  & $-$2.3 & 7.9 \\
IR4 & $-$10.9 & $-$10.1 & $-$10.2 & $-$10.3 & 133.2  & $-$2.3 & 7.9 \\ \hline
\end{tabular}
\end{table}

\begin{figure}
\resizebox{\hsize}{!}{\includegraphics{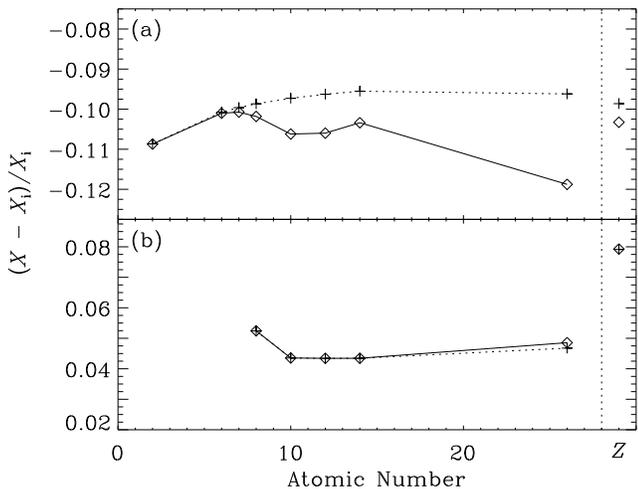}}
\caption{\mybf{\textbf{(a)} The abundance changes at the surface
between initial and solar model of sequence IR3 (dashed line) and IR4
(solid). On the right hand side the change in the total metallicity
$Z$ is shown. \textbf{(b)} As (a) but in the centre. All
elements lighter than \element[][]{O} have been omitted, as their
abundance is basically determined by the nuclear equilibrium of the
CNO cycle.}
\label{abund}}
\end{figure}

\begin{figure}
\resizebox{\hsize}{!}{\includegraphics{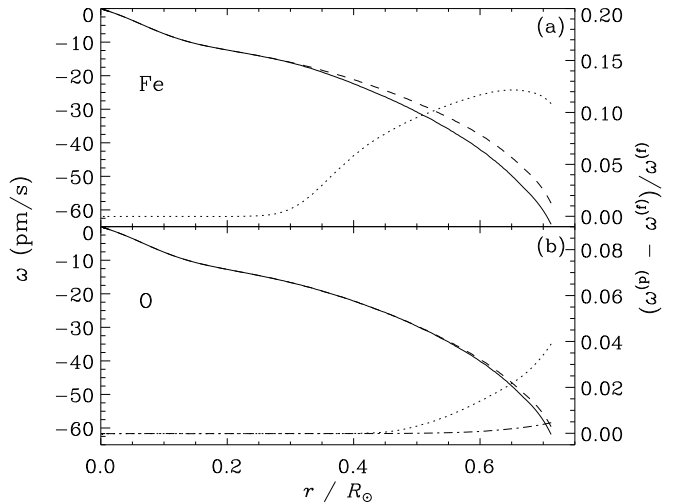}}
\caption{\mybf{\textbf{(a)} The diffusion velocity of \element[][]{Fe}
in the Sun assuming fully (dashed line) resp. partly (solid) 
ionized elements and their relative difference (dashed
line). \textbf{(b)} Like (a) but for \element[][]{O}. In addition the
relative difference in the diffusion velocity of \element[][]{C}
between part and complete ionization are shown, too (dash-dotted line).
}
\label{diffpartw}}
\end{figure}

The abundance variations of \element[][4]{He}, \element[][12]{C}, and
\element[][16]{O} at the solar surface and in the centre compared to
their initial values are summarized in Table~\ref{abunsol}. 
Considering the detailed ionization stages of the metals in the
diffusion scheme yield a higher reduction of \element[][16]{O} and
all metals on average, but the absolute changes remain small
($<0.4\%$). \mybf{The biggest change can be observed for iron, which
is about 2.5\% stronger depleted in model IR4 than in IR3
(Fig.~\ref{abund}). This is in agreement with the result of
\cite{TRM98}, but the overall change in 
the metal fraction of 0.4\% is about a factor of 2 less than found by
them.  The reason is a
smaller change of the diffusion velocity in
my computations by using partly instead of completely ionized metals.
In particular, the velocity of oxygen, which is the most abundant
element among the metals, is increased only by about 5\% below the convective
envelope (Fig.~\ref{diffpartw}), while \cite{TRM98} found an about 10\% higher value (see
Fig.~13, therein).}

\mybf{The cause of this difference is difficult to
disentangle, but it might be that the ionization fractions
obtained from the IRWIN-EOS do not agree with the ones of the
OPAL-EOS. The latter has been used by \cite{TRM98}.
A further clue supporting this assessment is that the surface
carbon abundance is almost unaltered in my models
(Table~\ref{abunsol}), while \cite{TRM98} found a depletion
of 8.5\% instead of 8\%. According to the IRWIN-EOS
the ionization of C below the solar convective envelope is nearly
complete~(Fig.~\ref{iondeg}a), i.e., 
no change in the carbon depletion is expected, if the improved
diffusion description is applied.}

Independent of whether complete or partial ionization is assumed in
the diffusion equation, the somewhat different diffusion velocities
of each element cause a slight redistribution of metals, such that
they are inconsistent with the metal composition assumed in the opacity
tables. \cite{TRM98} has shown that the usage of monochromatic
opacities may change the sound speed by about 0.1\% compared to the 
Sun. They obtained this difference in a model with elaborate
treatment of the ionization stages. However, as mentioned above, the
metal distribution in their model was modified stronger compared to
the initial one than in model IR4. So, using 
monochromatic opacities probably had a bigger effect on their model,
than it would have on model IR4.

\subsection{Low-mass stars} \label{lowmass}
In addition to the solar models also models for low-mass metal-poor
stars have been computed. \mybf{The aim is to examine} the influence of
the improved diffusion scheme on their TO properties. \mybf{Certainly
the inclusion of radiative levitation can alter the models considerably
\citep{RMR02}, but it is nevertheless important to show how the more
detailed treatment of ionization stages in the diffusion scheme
modifies the stellar models.}
 
Basically, the same
input physics has been used as for the solar models, where for the
equation of state \citeposs{Irwin} description has been applied. Since the
2D-atmospheres implemented in the solar models do not suffice in temperature,
gravity and abundances for metal-poor stars, the empirical
$T$-$\tau$-relation of \cite{Krish} has been included. For the convection 
\citeposs{MLT} mixing-length theory was employed with a mixing-length
parameter of 1.77, close to the calibrated value of a solar model
containing the same physics. The metal distributions in the models were
chosen to be $\alpha$-enhanced with
$\mathrm{[\alpha/Fe]=0.35}$. Consistent opacity tables have been produced
for this mixture \citetext{A. Weiss, private communication}.

\begin{table*}
\caption{Characteristic properties of low-mass stars at the TO with initial
$Z=10^{-3}$ and $Y = 0.24$, whereas an $\alpha$-enhanced mixture has 
been used with $\mathrm{[\alpha/Fe] = 0.35}$. $\Delta M_\mathrm{c.e.}$ is the mass of the   
convective envelope. To show the changes in the abundances $\{x\}
\equiv \log_{10}(x/x_i)$ has been defined, where the index $i$ denotes
the initial value. The first  
row for each mass represents the no-diffusion case, while the second
and third show the values considering fully and partly ionized elements,
respectively, in the diffusion equation.
\label{totab}} 
\begin{tabular}{l|ccccrrrrrrrrr} \hline \hline
$M_\odot$ & Age & $T_\mathrm{eff}~\mathrm{[K]}$ & $\frac{L}{L_\odot}$ &
$\frac{\Delta M_\mathrm{c.e.}}{M_{\odot}}$ &
\{$Y$\} & \{$Z$\} & \{$\frac{X_\mathrm{C}}{Z}$\} & \{$\frac{X_\mathrm{N}}{Z}$\} & 
\{$\frac{X_\mathrm{O}}{Z}$\} & \{$\frac{X_\mathrm{Ne}}{Z}$\} & \{$\frac{X_\mathrm{Mg}}{Z}$\} & 
\{$\frac{X_\mathrm{Si}}{Z}$\} & \{$\frac{X_\mathrm{Fe}}{Z}$\} \\ \hline 
     & 7.75 & 6760 & 2.68 & $9.5\times10^{-6}$ & 0.0 & 0.0 & 0.00  & 0.00  & 0.00  &
0.00    & 0.00    & 0.00    & 0.00\\
0.9  & 7.05 & 6613 & 2.40 & $7.7\times10^{-6}$ & $-$1.8 & $-$1.5 &
$-$0.06 & $-$0.03 & $-$0.01 & 0.03 & 0.05 & 0.07 & 0.07 \\
     & 7.07 & 6626 & 2.41 & $6.5\times10^{-6}$ & $-$1.9 & $-$2.1 &
$-$0.22 & $-$0.03 & 0.09 & $-$0.05 & $-$0.42 & $-$0.88 & $-$1.34 \\ \hline
     & 5.30 & 7248 & 4.03 & $6.0\times10^{-9}$ & 0.0 & 0.0     & 0.00    & 0.00    & 0.00    &
0.00    & 0.00    & 0.00    & 0.00\\
1.0  & 5.03 & 7117 & 3.86 & $5.6\times10^{-9}$ & $-$3.3 & $-$2.0 & $-$0.32 & $-$0.17 & $-$0.06 &
0.11 & 0.21 & 0.29 & 0.34 \\
     & 5.05 & 7154 & 3.89 & $4.0\times10^{-9}$ & $-$3.2 & $-$4.2 & $-$2.26 & $-$0.61 & 0.20 & 
$-$2.26 &  $-\infty$   & $-\infty$  & $-\infty$ \\ \hline
     & 3.63 & 7860 & 5.50  & $2.8\times10^{-9}$ & 0.0 & 0.0     & 0.00    & 0.00    & 0.00    &
0.00    & 0.00    & 0.00    & 0.00\\
1.1  & 3.48 & 7727 & 5.29  & $1.0\times10^{-10}$ & $-$2.6 & $-$1.4 &  $-$0.21 & $-$0.11 & 0.03 & 
0.08 & 0.14 & 0.19 & 0.23 \\
     & 3.50 & 7763 & 5.32  & $8.5\times10^{-11}$ & $-$2.5 & $-$2.8 &  $-$2.12 & $-$0.59 & 0.18 &
$-$0.32 & $-$5.34 & $-\infty$ & $-\infty$ \\ \hline
\end{tabular}
\end{table*}

\begin{figure}
\resizebox{\hsize}{!}{\includegraphics{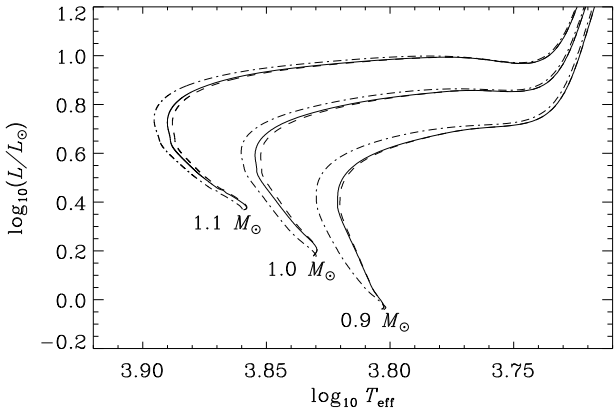}}
\caption{The evolution of 0.9, 1.0, and 1.1~$M_\odot$ stars with
$\mathrm{[\element[][]{Fe}/\element[][]{H}]=-1.6}$ ($\alpha$-enhanced) treating all elements in
the diffusion as partly (solid line) or fully ionized (dashed). For comparison the
evolutions without diffusion are shown, too (dash-dotted line).
\label{hrd_pop2}}
\end{figure}

In Fig.~\ref{hrd_pop2} the evolution of 0.9, 1.0, and 1.1~$M_\odot$
stars from the zero-age main sequence until the early-RGB are
displayed in the HR-diagram. Their properties at the TO are summarized in
Table~\ref{totab}. The inclusion of diffusion leads to smaller ages at
the TO, which has also been found by various authors. The effect of
decreasing the age is more pronounced for smaller masses which have 
longer main-sequence lifetimes. There, diffusion is capable of
decreasing the core hydrogen content considerably, the amount of which
basically determines the time spent on the main sequence.

Strongly varying pressure and temperature gradients near the surface
of these stars cause an almost total depletion of helium and metals at the
surface near the TO phase. Thereby the degree of the helium depletion
depends on two parameters, the extension of the convective envelope,
which provides the reservoir for inward migrating atoms, and the time
the star needs to reach the TO, i.e., the time diffusion can
operate. The competing effects of these two parameters can be observed in the
evaluated models. Since the mass of the convective envelope of the
1.0~$M_\odot$ star is orders of magnitude smaller than in the
0.9~$M_\odot$ star, helium is much more
depleted in the former than in the latter case (Table~\ref{totab}).
However, although the extension of the convective envelope is smallest
in the 1.1~$M_\odot$ star, more \He~remains on the surface than in the 
1.0~$M_\odot$ case; the reason for this is the smaller lifetime of
the 1.1~$M_\odot$ star on the main sequence. 

The surface metal abundances follow the trend of He. However, their
total depletion and distribution is strongly depending on the
treatment of ions in the diffusion equation. Assuming complete ionization
leads in general to smaller diffusion velocities (cf.~Fig~\ref{feion}) and 
thus to a smaller decrease of the surface metal content. Already in
this case the metal distribution may be altered up to a
factor of 2 (Table~\ref{totab}). Taking into account the different
ionization stages some metals may even totally disappear from the surface.

\begin{figure}
\resizebox{\hsize}{!}{\includegraphics{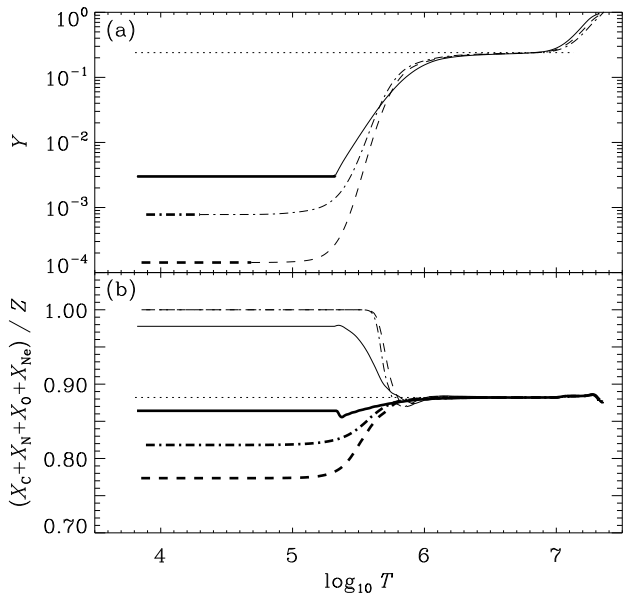}}
\caption{\textbf{a)} The helium profile of a 0.9 (solid line), a 1.0 (dash-dotted), and
a 1.1 (dashed) $M_\odot$ star at the TO including diffusion with
elaborate treatment of the ionization stages. The thicker 
lines mark the envelope convective zones. \textbf{b)} The
contribution of \element[][]{C}, \element[][]{N}, \element[][]{O}, and
\element[][]{Ne} to the total metallicity. The
line styles agree with a), however, the thick lines correspond to
models with complete ionization assumed. The dotted lines represent in both
panels the initial values.\label{surf}}
\end{figure}

However, whether
this also occurs in real stars is questionable, as with decreasing
abundance and increasing effective temperature the influence of radiative
levitation, which has been neglected here, becomes stronger. 
It would diminish the tendency of metals to settle in the
stellar interior and might even lead to an enhancement of some elements
on the surface \citep{RMR02}. 

An additional source of uncertainty in these models is given by the
strong deviation of the metal distribution from the initial one, being
apparent in particular in models with detailed ionization treatment. With an assumed standard $\alpha$-enhanced mixture
in the opacity tables, the opacities obtained are not valid
any more at the TO, which may lead to different effective
temperatures. However, \cite{MCS93} have demonstrated that the TO temperature
of a $M=0.8\,M_\odot$ star with a metallicity $Z=10^{-3}$ is
determined basically by the opacities above about 12,000\,K. Moreover,
the opacities there are only sensitive to the contribution of C,
N, O, and Ne to the metals, whereas the detailed distribution of these
elements is only of minor importance. Artificially decreasing their
abundances, i.e., 
underestimating their contribution, yields too high effective
temperatures.

In the models of this work, where full ionization is assumed, the
contribution of C, 
N, O, and Ne has been overestimated using standard $\alpha$-enhanced
opacity tables (Fig.~\ref{surf}b), implying that the obtained effective
temperature are actually too small. In way of contrast, just the
opposite behaviour can be observed in the models where the ionization
stages are considered in detail. The result is that the actual
difference in effective temperature between models assuming either
complete or partial ionization would be smaller as quoted in
Table~\ref{totab}. From the work of \cite{MCS93}, I conclude
that the reduction would be about 10\,K.

Thus, when the redistributed metal abundances are taken into account
in the opacities, the effective temperature of the 1.1~$M_\odot$ star
is increased only be about 20~K by treating ionization in 
the diffusion equation in detail. The effective temperature of the
0.9~$M_\odot$ star
remains almost unaltered. 
Probably, radiative levitation may slightly alter the effective
temperature, but it is not expected that it alters the presented
results by much more than 10~K. In fact, \cite{RMR02} have found an 
increase of about 10\,K for a 0.8~$M_\odot$ star when including
radiative levitation and using monochromatic opacities.
For the age determination of globular clusters the exact diffusion
treatment is therefore only of minor
importance, in
particular, as the effect of the improved diffusion treatment
is waning with decreasing TO mass, and thus increasing
globular-cluster age.   

\section{Summary}
Starting from Burgers' equation for a multicomponent fluid the effect of
partly ionized metals has been consistently included in the diffusion
equation following the procedure of TBL. As an additional feature,
electron degeneracy and the contribution of radiation to the total
pressure has been taken into account. 
It has been shown, that for most stars it is sufficient to use one ion
per element carrying the mean charge instead of computing diffusion
for each ion separately. By this means, the elaborate treatment of
partly ionized metals remains feasible. 
With increasing stellar mass,
and thus decreasing stellar convective envelope,  it may become
necessary to consider, at least, all \He-ions separately.
However, since these stars are usually hotter, radiative levitation
becomes important, which then should be included, too. 

The effect of the improved diffusion treatment of partly ionized
metals has been investigated in case of the Sun and the TO properties of
low-mass metal-poor stars. In order to obtain the accurate ionization
degrees of each metal species the equation of state of \cite{Irwin} has been
employed. The sound-speed profile in solar models obtained with this
new EOS is in as good agreement with the helioseismic determined one
as models with the updated OPAL01-EOS.

An improvement of up to 25\% between $r=0.25 R_\odot$ and $r=0.65
R_\odot$ could be achieved by including the effect of partly 
ionization into the diffusion equation. However, this value is
presently still exceeded by the uncertainties in other input physics
like the EOS. 
Nevertheless, the changes between models treating metals
fully and partly ionized have been found to be smaller
than claimed by \cite{TRM98}. \mybf{The origin of this discrepancy is
not clear, but might be due to the use of different equations of state.}

In low-mass metal-poor stars the improved diffusion description causes
strong deviations in the surface metal distribution from the
initial one. In addition, the depletion of metals is much stronger than
in the case when full ionization is assumed. The consequent
change in the effective temperature is at most about 40~K for a
1.1~$M_\odot$ star diminishing
with decreasing mass. 

The influence of radiative levitation on the diffusive
behaviour in particular of the metals has been neglected in this work.  
This may reduce the amount of depletion, and may further alter the
metal distribution. Besides, since the effect of using partly instead
of fully ionized metals in the diffusion equation is biggest for small
convective envelopes, a small amount of additional mixing
below the convective boundary may reduce the effect of diffusion
strongly. 

With the mass of the convective envelopes of metal-poor
stars with 
$M>1~M_\odot$ being very small ($\la 10^{-8}~M_\odot$), also the
amount of mass loss may strongly modify the surface abundances.
However, mass-loss mechanisms are only very poorly understood. 
Therefore, first of all, precise stellar models are needed to
disentangle the extent of mass loss and to determine the influence of
rotation. 

\acknowledgement{This work has been supported by a Marie Curie
Fellowship of the European Community programme ``Human Potential''
under contract number HPMF-CT-2000-00951. I would like to thank
M.\  Salaris for useful comments and inspiring discussions, and
A.\  Weiss, who kindly provided splined versions of the opacity
tables. In addition, I am grateful to A.W.\  Irwin for providing me his
equation of state.}  

\appendix
\section{}\label{appx}
The algebraic transformations of Eqs.~\bref{temp}--\bref{localmass} are
toilsome but mathematically not very demanding. Therefore, only
the basic intermediate steps and the final results are described here.

First, the r.h.s. of Eqs.~\bref{press} is converted, which consists
according to Eq.~\bref{psequ} of two terms, where the first one, 
$\frac{\tot \ln\beta P}{\tot r}$, is almost in its final shape
\[
\frac{\tot \ln\beta P}{\tot r} = \frac{\tot\ln P}{\tot r} + \frac{\tot
\ln \beta}{\tot r}.
\]
Using Eq.~\bref{neutc}, the remaining one can be written as

\begin{eqnarray}\label{cfunc}
\frac{\tot \ln (C_s/C)}{\tot r}\! & = & \!
\sum_{t\neq \He,e}\left(\delta_{ts} - \frac{C_t}{C}\right) \frac{\tot
\ln C_t}{\tot r}- \left(\delta_{s,\He} - \frac{C_{\He}}{C}\right)
\times \nonumber \\
&& \hspace{-1cm}\left(\sum_{t\neq\He} \frac{\zeta_t
\tilde{C}_t}{\zeta_{\He} C_{\He}} \frac{\tot
\ln \tilde{C}_t}{\tot r} +  \sum_{t\neq e}\frac{\zeta_t
C_t}{\zeta_{\He} C_{\He}} \frac{\tot 
\ln \zeta_t}{\tot r}\right), \\ \nonumber
\end{eqnarray}
where ``$\He$'' denotes either of the helium ions (not the neutral He!), and 
\[
\tilde{C}_s = \left\{ \begin{array}{l}
C_s,\quad\mbox{for all ions} \\
\tilde{C}_e,\quad\mbox{for electrons.}
\end{array}\right. 
\]
The l.h.s. of the temperature
equation (Eq.~\bref{temp}) is given by Eq.~\bref{tfunc}.

The equations for current neutrality (Eq.~\bref{neutcur}) and local
mass conservation (Eq.~\bref{localmass}) become, using Eq.~\bref{defc},
\begin{equation}\label{afunc}
\sum A_s \tilde{C}_s \omega_s = 0
\end{equation}
and
\begin{equation}\label{zfunc}
\sum \zeta_s \tilde{C}_s \omega_s = 0,
\end{equation}
where $A_s$ is the mass of ion $s$ in atomic units.
The system of equations~\bref{temp}--\bref{localmass} can now be
written in the form

\begin{eqnarray*}
\lefteqn{\frac{\beta P}{K_0}\biggl[\alpha_i\frac{\tot \ln P}{\tot r} +
\varphi_i\frac{\tot \ln \beta}{\tot r} + \nu_i \frac{\tot \ln
T}{\tot r} +\sum_{j=1 \atop j\neq 2}^S \gamma_{ij}\frac{\tot \ln \tilde{C}_j}{\tot r}
} && \\
&\hspace{3cm} & + \sum_{j=1 \atop j\neq S}^S \lambda_{ij} \frac{\tot \ln \zeta_j}{\tot
r}\biggr] = \sum_{j=1}^{2S+2} \Delta_{ij} W_j,\\
\end{eqnarray*}
where $\He$ is supposed to be species No.~2 and the electrons species
No.~$S$. The quantities $K_0$, $\vec{\alpha}$, $\vec{\nu}$,
$\tens{\gamma}$, $\tens{\Delta}$, and $\vec{W}$ are defined as in
TBL with the following exceptions:
\begin{enumerate}
\item{$\gamma_{ij} = 
\frac{C_i}{C}\left[(1 - \delta_{jS})\left(\delta_{ij} -
\frac{C_j}{C}\right) - \left(\delta_{i2} - \frac{C_2}{C}\right)\frac{\zeta_j
\tilde{C}_j}{\zeta_2 C_2}\right]$ for \\ \mbox{$i=1,\dots,S$}} (cf.~Eq.~\bref{cfunc})
\item{$\Delta_{ij} = \left\{\begin{array}{lll} 
A_j \tilde{C}_j & \mathrm{for}~j=1,\dots,S; i=2S+1 &
\mathrm{(cf.~Eq.~\bref{afunc})} \\
\zeta_j \tilde{C}_j & \mathrm{for}~j=1,\dots,S; i=2S+2 &
\mathrm{(cf.~Eq.~\bref{zfunc})} \\
\end{array}\right.$}
\item{$\Delta_{ij} = \left\{\begin{array}{ll}
\zeta_j \tilde{C}_j & \mathrm{for}~j=2S+1; i=1,\dots,S \\
-A_j \tilde{C}_j & \mathrm{for}~j=2S+2; i=1,\dots,S\\
\end{array}\right.$}
\item{$W_j = \left\{\begin{array}{ll}
K_0^{-1}\tilde{n}_e e E & \mathrm{for}~j=2S+1\\
K_0^{-1}\beta\tilde{n}_e m_0 g & \mathrm{for}~j=2S+2,\\
\end{array}\right.$}
\end{enumerate}
where $m_0$ is the atomic mass unit and $K_0 =
\frac{3}{4}\sqrt{\frac{m_0}{2\pi}}k_\mathrm{B}^{5/2}e^{-4}$.
Finally, the new coefficients $\vec{\varphi}$ and $\tens{\lambda}$ are defined (see Eq.~\bref{cfunc})
as
\[
\varphi_i = \frac{C_i}{C} - \frac{A_i \tilde{C}_i}{\sum_i A_i \tilde{C}_i}\quad\mathrm{for}~i=1,\dots,S
\]
and
\[
\lambda_{ij}=\left\{\begin{array}{ll}
-\frac{C_i}{C}\left(\delta_{i2} -\frac{C_2}{C}\right)\frac{\zeta_j
C_j}{\zeta_2 C_2} & \mathrm{for}~i=1,\dots,S-1 \\
0 &  \mathrm{for}~i=S,\dots,2S+2.
\end{array}\right.
\]
To account for the partly ionized metals in the particle interaction,
TBL's $Z_s$ has to be substituted by $\zeta_s$ in the cross section
$\sigma_{st}$ (TBL's Eq.~(8)), and in the Coulomb
logarithm $\ln \Lambda_{st}$ (TBL's Eq.~(9)). 


\begin{thebibliography}{36}
\expandafter\ifx\csname natexlab\endcsname\relax\def\natexlab#1{#1}\fi

\bibitem[{Adelberger {et~al.}(1998)Adelberger, Austin, Bahcall, Balantekin,
  Bogaert, \& Brown}]{Adel98}
Adelberger, E.~G., Austin, S.~M., Bahcall, J.~N., {et~al.} 1998, \rmp, 70, 1265

\bibitem[{Babel \& Michaud(1991)}]{BM91}
Babel, J. \& Michaud, G. 1991, \aa_, 248, 155

\bibitem[{Bahcall {et~al.}(1998)Bahcall, Basu, \& Pinsonneault}]{Bah98}
Bahcall, J.~N., Basu, S., \& Pinsonneault, M.~H. 1998, \plb, 433, 1

\bibitem[{Bahcall \& Pinsonneault(1992)}]{BP92}
Bahcall, J.~N. \& Pinsonneault, M.~H. 1992, \rmp, 64, 885

\bibitem[{Basu {et~al.}(1997)Basu, Chaplin, Christensen-Dalsgaard, Elsworth,
  Isaak, New, Schou, Thompson, \& Tomczyk}]{BCC97}
Basu, S., Chaplin, W.~J., Christensen-Dalsgaard, J., {et~al.} 1997, \mnras,
  292, 243

\bibitem[{B{\"o}hm-Vitense(1958)}]{MLT}
B{\"o}hm-Vitense, E. 1958, \zap, 46, 108

\bibitem[{Bonanno {et~al.}(2001)Bonanno, Murabito, \& Patern{\`o}}]{BMP01}
Bonanno, A., Murabito, A.~L., \& Patern{\`o}, L. 2001, \aa_, 375, 1062

\bibitem[{Bonanno {et~al.}(2002)Bonanno, Schlattl, \& Patern{\`o}}]{BSP02}
Bonanno, A., Schlattl, H., \& Patern{\`o}, L. 2002, \aa_, 390, 1115

\bibitem[{Burgers(1969)}]{Burgers}
Burgers, J.~M. 1969, Flow Equations for Composite Gases (New York, London:
  Academic Press)

\bibitem[{Canuto \& Mazzitelli(1992)}]{CM2}
Canuto, M. \& Mazzitelli, I. 1992, \apj, 389, 724

\bibitem[{Castellani {et~al.}(1997)Castellani, Ciacio, degl'Innocenti, \&
  Fiorentini}]{CCIF97}
Castellani, V., Ciacio, F., degl'Innocenti, S., \& Fiorentini, G. 1997, \aa_,
  322, 801

\bibitem[{Chaboyer {et~al.}(2001)Chaboyer, Fenton, Nelan, Patnaude, \&
  Simon}]{CFN01}
Chaboyer, B., Fenton, W.~H., Nelan, J.~E., Patnaude, D.~J., \& Simon, F.~E.
  2001, \apj, 562, 521

\bibitem[{Eggleton {et~al.}(1973)Eggleton, Faulkner, \& Flannery}]{EFF}
Eggleton, P.~P., Faulkner, J., \& Flannery, B.~P. 1973, \aa_, 23, 325

\bibitem[{Freytag {et~al.}(1996)Freytag, Ludwig, \& Steffen}]{2dhydro}
Freytag, B., Ludwig, {\singlet{H.-G}}., \& Steffen, M. 1996, \aa_, 313, 497

\bibitem[{Gonzalez {et~al.}(1995)Gonzalez, LeBlanc, Artru, \& Michaud}]{GLAM95}
Gonzalez, {\singlet{J.-F}}., LeBlanc, F., Artru, {\singlet{M.-C}}., \& Michaud,
  G. 1995, \aa_, 297, 223

\bibitem[{Grevesse \& Noels(1993)}]{GN93}
Grevesse, N. \& Noels, A. 1993, Phys.~Scripta, T47, 133

\bibitem[{Iglesias \& Rogers(1996)}]{OP96}
Iglesias, C.~A. \& Rogers, F.~J. 1996, \apj, 464, 943

\bibitem[{Irwin(2000)}]{Irwin}
Irwin, A.~W. 2000, GNU public license, {FORTRAN} routines available at\ {\tt
  ftp://astroftp.phys.uvic.ca/pub/irwin/eos/code}

\bibitem[{Kippenhahn {et~al.}(1967)Kippenhahn, Weigert, \& Hofmeister}]{KWH}
Kippenhahn, R., Weigert, A., \& Hofmeister, E. 1967, Methods in Computational
  Physics, Vol.~7, Methods for Calculating Stellar Evolution (New York:
  Academic Press), 129

\bibitem[{\mbox{Krishna Swamy}(1966)}]{Krish}
\mbox{Krishna Swamy}, K.~S. 1966, \apj, 145

\bibitem[{Michaud {et~al.}(1978)Michaud, Martel, \& Ratel}]{MMR78}
Michaud, G., Martel, A., \& Ratel, A. 1978, \apj, 226, 483

\bibitem[{Ram{\'\i}rez \& Cohen(2002)}]{RC02}
Ram{\'\i}rez, S.~V. \& Cohen, J.~G. 2002, \aj, 123, 3277

\bibitem[{Ram{\'\i}rez {et~al.}(2001)Ram{\'\i}rez, Cohen, Buss, \&
  Briley}]{RCBB01}
Ram{\'\i}rez, S.~V., Cohen, J.~G., Buss, J., \& Briley, M.~M. 2001, \aj, 122,
  1429

\bibitem[{Richard {et~al.}(2002)Richard, Michaud, Richer, Turcotte,
  Turck-Chi{\`e}ze, \& VandenBerg}]{RMR02}
Richard, O., Michaud, G., Richer, J., {et~al.} 2002, \apj, 568, 979

\bibitem[{Richard {et~al.}(1996)Richard, Vauclair, Charbonnel, \&
  Dziembowski}]{Richard96}
Richard, O., Vauclair, S., Charbonnel, C., \& Dziembowski, W.~A. 1996, \aa_,
  312, 1000

\bibitem[{Richer {et~al.}(2000)Richer, Michaud, \& Turcotte}]{RMT00}
Richer, J., Michaud, G., \& Turcotte, S. 2000, \apj, 529, 338

\bibitem[{Rogers(2001)}]{Rog01}
Rogers, F.~J. 2001, \cpp, 41, 179

\bibitem[{Rogers {et~al.}(1996)Rogers, Swenson, \& Iglesias}]{OPEOS}
Rogers, F.~J., Swenson, F.~J., \& Iglesias, C.~A. 1996, \apj, 456, 902

\bibitem[{Salaris {et~al.}(1993)Salaris, Chieffi, \& Straniero}]{MCS93}
Salaris, M., Chieffi, A., \& Straniero, O. 1993, \apj, 414, 580

\bibitem[{Salaris {et~al.}(2000)Salaris, Groenewegen, \& Weiss}]{SGW00}
Salaris, M., Groenewegen, M. A.~T., \& Weiss, A. 2000, \aa_, 355, 299

\bibitem[{Schlattl(1999)}]{PhD}
Schlattl, H. 1999, \mbox{Ph.D.~thesis}, Technical University Munich

\bibitem[{Schlattl(2001)}]{S00}
---. 2001, \prd, 64, 013009

\bibitem[{Schlattl {et~al.}(1997)Schlattl, Weiss, \& Ludwig}]{SWL97}
Schlattl, H., Weiss, A., \& Ludwig, {\singlet{H.-G}}. 1997, \aa_, 322, 646

\bibitem[{Thoul {et~al.}(1994)Thoul, Bahcall, \& Loeb}]{Diffc}
Thoul, A.~A., Bahcall, J.~N., \& Loeb, A. 1994, \apj, 421, 828

\bibitem[{Turcotte {et~al.}(1998)Turcotte, Richer, Michaud, Iglesias, \&
  Rogers}]{TRM98}
Turcotte, S., Richer, J., Michaud, G., Iglesias, C.~A., \& Rogers, F.~J. 1998,
  \apj, 504, 539

\bibitem[{Weiss \& Schlattl(2000)}]{WS00}
Weiss, A. \& Schlattl, H. 2000, \aas, 144, 487

\end{thebibliography}

\newcommand{\singlet}[1]{#1}

\end{document}